\documentclass[twocolumn,aps,pre,tightenlines,superscriptaddress,nofootinbib]{revtex4-1}
\usepackage{amsmath,amssymb,amsfonts,graphicx,times}

\usepackage[all]{xy}
\usepackage{hyperref}
\usepackage[usenames,dvipsnames]{color}
\definecolor{darkred}{rgb}{0.8,0.1,0.1}
\hypersetup{colorlinks=true, linkcolor=darkred, citecolor=blue}

\begin{document}
\title{Quench echo and work statistics in integrable quantum field theories}

\author{T. P\'almai}
\email{tpalmai@sissa.it}

\affiliation{\textit{\small SISSA and INFN, Sezione Trieste, Via Bonomea 265, 34136 Trieste, Italy}{\small{} }}
\affiliation{\textit{\small Budapest University of Technology and Economics, Budafoki ut 8, 1111 Budapest, Hungary}{\small{} }}

\author{S. Sotiriadis}
\email{sotiriadis@df.unipi.it}

\affiliation{\textit{\small Dipartimento di Fisica dell'Universit\`a di Pisa and INFN, Sezione Pisa, 56127 Pisa, Italy}{\small{} }}
\affiliation{\textit{\small SISSA and INFN, Sezione Trieste, Via Bonomea 265, 34136 Trieste, Italy}{\small{} }}

\begin{abstract}
We propose a boundary thermodynamic Bethe ansatz calculation technique to obtain the Loschmidt echo and the statistics of the work done when a global quantum quench is performed on an integrable quantum field theory. We derive an analytic expression for the lowest edge of the probability density function and find that it exhibits universal features, in the sense that its scaling form depends only on the statistics of excitations. We perform numerical calculations on the sinh-Gordon model, a deformation of the free boson theory, and we obtain that by turning on the interaction the density function develops fermionic properties. The calculations are facilitated by a previously unnoticed property of the thermodynamic Bethe ansatz construction.

\end{abstract}

\maketitle

\section{Introduction}

Understanding the statistical properties of quantum systems out of
equilibrium is one of the main challenges of modern physics. Apart
from numerous applications to a diversity of physical problems (e.g.
inflationary expansion of the early universe in cosmology, large-scale
quantum computation) and experiments (e.g. heavy-ion collision, dynamics
of cold atomic gases), this branch of physics is inherently related
to a fundamental long-standing question: How and under what conditions
a quantum system, that is initially prepared in an out-of-equilibrium state and evolves under the quantum mechanical law of unitary evolution,  tends for long times to equilibrium \cite{d91,sr94,Rigol08}? Is this equilibrium always thermal or of some generalized type \cite{GGE,GGEr}? 
Experimental progress over the last decade \cite{gmhb-02,kww-06,tetal-11,expR}
followed by advances in the theoretical treatment of out-of-equilibrium
quantum physics have opened a unique opportunity to tackle
such questions. A special protocol that has concentrated
a great part of attention for its relatively simple theoretical treatment
and experimental feasibility is the \emph{quantum quench} protocol \cite{cc-06,cc-07,c-06,caz1,caz2,caz3,r-09,r-09a,scc-09,bs-08,CE-08,CE-10,BKL-10,CCR-11,bdkm-12,sks-13,ekmr-14,r-14,wrdk-14,Mossel,CEF,CEF-i,f-13,eef-12,se-12,ccss-11,rs-12,SE,CSC13a,CSC13b,KCC-13,f-14,ck-14,ce-13,RS13,FM,Pozsgay-11,SFM,mc-12b,dt-12,ck-12,a1,a2,a3,a4,M13,gge-new-1,gge-new-2,gge-new-3,gge-new-5,pos-13,STM,fcec-13,bdwc-14,b-14,cd-14,SC-14}: an 
instantaneous change of parameters of the Hamiltonian of a system
so that it initially lies in the ground state of the pre-quench
Hamiltonian while it evolves under the post-quench Hamiltonian. 

An important quantity in the study of quantum quench problems is the
following one
\begin{equation}
Z(z)=\langle\Omega|e^{-Hz}|\Omega\rangle\label{eq:Z}
\end{equation}
where $|\Omega\rangle$ is the initial state (or boundary-in-time
state), i.e. the ground state of the pre-quench Hamiltonian, and $H$
the post-quench Hamiltonian. For imaginary values $z=it$ this function
gives the overlap between the initial and the evolved state after
time $t$, the Loschmidt amplitude, whose norm square is the \emph{Loschmidt echo} $L(t)$ \cite{LE01,LE02,LE03,LE04,LE05}. (In the more general case when $|\Omega\rangle$ is not an eigenstate of the pre-quench Hamiltonian, the Loschmidt echo is modified by a nontrivial evolution under the prequench Hamiltonian $H_0$, see e.g. \cite{LE03}.) This overlap is essentially the characteristic function of the probability density function $P(W)$ of the statistics of the work done \cite{FTHEO,RMPQS,S1,GS1,SGS} (differing in a constant factor and shift). Knowledge of this distribution amounts to knowledge of all probabilities to excite each post-quench energy level, which is all that is needed in order to calculate the long time averages of observables \cite{M13} or their asymptotic values provided they become stationary \cite{FM}. On the other hand, for positive real values of $z=R$ the above quantity is the partition function of a system defined in a 2d strip of width $R$ with boundary conditions $|\Omega\rangle$ on 
both edges. Thus $Z(z)$ provides a direct manifestation of a useful mapping between the out-of-equilibrium problem of a quantum quench and a well-studied boundary problem \cite{cc-06,cc-07,GS1}.

$Z(z)$ in the complex plane is also an interesting subject of study. In lattice models in particular, since $Z(z)$ is an entire function \cite{Fisher}, its zeroes in the complex $z$-plane determine completely the non-analytic part of the associated free energy. It has been argued that, in the thermodynamic limit a concentration of zeroes around particular points of the imaginary axis (real $t$-axis) is associated to non-analytic behaviour in the free energy \cite{HPK}. This has been demonstrated for a quench across the critical point of the Ising model, exploiting its mapping to free fermions (where however the continuum limit tames the singularities of the free energy corresponding to zeroes of $Z(z)$ into square root branch points). It was proposed that such zeroes (or branch points) of $Z(z)$ play a role analogous to that of Fisher zeroes in ordinary phase transitions \cite{HPK}. This non-analytic behaviour was later shown to be robust under the inclusion of non-integrable interactions, irrelevant or 
relevant in the RG sense, using the tDMRG numerical method \cite{KS}. However more recently it was also shown that such non-analyticities are caused more generally by a crossing of eigenvalues in the spectrum of the transfer matrix \cite{AS} and that they may not appear even if a critical line is crossed, thus indicating that the presence of zeros in $Z(z)$ and non-analyticities in the free energy and $L(t)$ are not a characteristic feature of dynamical phase transitions \cite{vd-14}.
Further progress in the analytical calculation of the above quantity
for integrable spin chains was made using Algebraic Bethe Ansatz techniques
\cite{F13,P13}. If the analytical continuation of the complex variable
$z=R+it$ from real to imaginary values is not prevented by the presence
of zeroes (that correspond to logarithmic singularities of the associated
free energy), it is possible to calculate the work probability distribution
$P(W)$ from the boundary partition function $Z(R)$. In this way
universal behaviour associated with the critical Casimir effect in
the boundary formulation of the problem can be connected to universal
properties of $P(W)$ in the limit of massless post-quench Hamiltonian,
in particular properties of its lowest excitation threshold \cite{S1,SGS,GS1}.
Study of the return amplitude in a finite system described by a Conformal Field Theory reveals that in a certain time-to-system-size limit the reduced density matrix of the system approaches exponentially a thermal one, as well as the complicate structure of quantum revivals \cite{c-14}.
The Loschmidt echo has been calculated in the context of quantum quench
and related problems in several other studies relative to spin chains, conformal field theories and other models that are essentially equivalent to non-interacting ones \cite{S2,GS2,S3,S4,S5,LE1,LE2,LE3,LE4,LE5,LE6,LE7,LE8,LE9,LE10}. 

So far no analytical study of the Loschmidt echo has been done in an Integrable Quantum Field Theory (IQFT) that cannot be mapped into a non-interacting model (apart from a proposed generalization of a calculation based on Conformal Field Theory for the case of a local quantum quench problem \cite{LE11}).
For global quantum quenches in such a general relativistic IQFT and under certain conditions for the initial state, the function (\ref{eq:Z}), viewed as the partition function of the strip boundary problem for real values of $z$, can be calculated 
using the so-called \emph{boundary thermodynamic Bethe ansatz} (BTBA) \cite{TBA,BTBA,LMSS}. In the BTBA the boundaries are represented by states that preserve the integrability of the bulk of the system (\emph{boundary integrable states}) and were shown to be of the following form \cite{GZ}
\begin{equation}
\exp\left(\int_{0}^{\infty}\frac{d\theta}{2\pi}K(\theta)Z^{\dagger}(-\theta)Z^{\dagger}(\theta)\right)|0\rangle\label{eq:BS}
\end{equation}
The operators $Z(\theta),Z^\dagger(\theta)$ are the Zamolodchikov-Faddeev operators \cite{ZF1,ZF2}, strings of which acting on the vacuum create asymptotic scattering states, that constitute an eigenstate basis of the IQFT. $K(\theta)$ in the boundary integrable problem satisfies the so-called boundary cross-unitarity condition and is related to the reflection matrix ($\theta$ is the rapidity variable, a convenient reparametrization of the momentum of single particle excitations, i.e. its energy's and momentum's being $E=m\cosh\theta$ and $p=m\sinh\theta$, respectively). For simplicity, in the above equation and in all that follows we restrict ourselves to an IQFT with a single particle species, as is the sinh-Gordon (shG) model. The Dirichlet state, a typical example of a boundary integrable state, defined as the state annihilated by the physical field $\phi$ and thus having vanishing field fluctuations, is of the above form
with $K(\theta)=K_{D}(\theta)$ known exactly \cite{G,CT}.

The initial state of a \emph{global} quantum quench in IQFTs is given, in certain important special cases, by this same boundary integrable form (\ref{eq:BS}). Such cases are those of free theories and when the pre-quench mass tends to infinity. In the latter case the field fluctuations vanish and therefore the initial state is exactly the Dirichlet state. This state requires careful ultraviolet regularization, since otherwise it leads to divergent physical observables. It has been recently argued \cite{STM} that for a quantum quench in the shG model from a large but finite pre-quench mass $m_{0}$ and zero interaction to an arbitrary post-quench mass $m$ and arbitrary interaction, one obtains as initial state a modified Dirichlet state that is approximately given by the same form (\ref{eq:BS}) with amplitude $K(\theta)= -K_{D}(\theta)K_{FB}(\theta)$ where $K_{FB}(\theta)$ is the amplitude that corresponds to a free bosonic quantum quench of the mass from $m_{0}$ to $m$. This state is free from ultraviolet 
divergences and consistent with the requirements of a quantum quench.
In complete generality, the situation seems to be more complicated \cite{SFM}: the extensivity of the conserved charges in the initial state introduces certain constraints for the amplitudes of excitations contained in it, at least in the limit of large number of particles \cite{M13}, but those constraints are more general than the precise special form (\ref{eq:BS}).

Considering all the above, in this work we propose a derivation of the Loschmidt amplitude and thus the work statistics relative to a global quench in IQFTs by means of the boundary TBA solved on the imaginary axis. Since the TBA is an integral equation representation, we a priori expect that the analytical continuation would give the correct Loschmidt echo. Indeed, a similar approach has been successfully
used before for the calculation of excitation energies in IQFT \cite{DT1,DT2,DT3,BLZ97}.
We perform this calculation in the sinh-Gordon model, arguably the simplest relativistic IQFT with non-trivial S-matrix. While this serves as a demonstration of the approach, it also allows us to see the effects of turning on the interaction on the thermodynamics of a quench in a quantum field theory. On the other hand, it can be considered as the first step for the application to a quench in the sine-Gordon model, which can be obtained by analytical continuation of the interaction coupling parameter of the shG model from real to imaginary values. A quantum quench in the sine-Gordon model can actually be experimentally implemented, for example in systems of split 1d Bose-Einstein quasi-condensates that interact through a longitudinal potential barrier that behaves like a Josephson junction \cite{G1,G2}.

In what follows we expose general properties about the statistics of the work done when performing a quantum quench, then we introduce the thermodynamic Bethe ansatz, followed by a discussion of its continuation to imaginary temperatures, both from a theoretical and a numerical point of view. Then we discuss results for the work statistics of integrable field theories, in particular the sinh-Gordon model. We find that the first peak in the probability density function is not affected by the interaction and we give it in an analytic form. Then we discuss global mass quenches from large but finite initial masses in the free bosonic, free fermionic and the sinh-Gordon models. We determine that while the moments of the distribution, in particular the mean, are only changed slightly by turning on the interaction, the details are modified. This can be seen on the first edge, which develops a fermionic, positive edge singularity exponent opposed to the bosonic negative one. The last section is reserved for 
conclusions.

\section{Formulation}

\subsection{Work statistics of a quantum quench}

Consider a closed quantum system in a $d$-dimensional box of edge length
$L$ ($d$ is the number of spatial dimensions) with periodic boundary
conditions, undergoing a quench from the Hamiltonian $H_{0}$ to $H$.
Before the quench the system lies in the ground state $|\Omega\rangle$
of the Hamiltonian $H_{0}$. One can treat the quench as a thermodynamic
transformation and ask how much work is done on the system. 
The performed work can be defined by referring to two projective energy measurements before
and after the quench. Since the quench protocol is abrupt, the work becomes stochastic with a probability distribution

\begin{equation}\label{defPW}
P(W)=\sum_{\text{all eigenstates }\left|\Psi\right\rangle }\delta(W-E_{\Psi}+E_{gs,0})\left|\left\langle \Psi|\Omega\right\rangle \right|^{2}
\end{equation}
where $|\Psi\rangle$ is any eigenstate of the post-quench Hamiltonian,
$E_{\Psi}$ the corresponding energy eigenvalue and $E_{gs,0}$ the
energy of the pre-quench ground state $|\Omega\rangle$. The characteristic
function $G(t)$ of this probability distribution is the inverse
Fourier transform 
\begin{eqnarray}
G(t) & = & \int_{-\infty}^{\infty}dW\ e^{-iWt}P(W)
\end{eqnarray}
which, using the previous definition, can be readily shown to be equal
to 
\begin{eqnarray}
G(t) & = & \sum_{\left|\Psi\right\rangle }e^{i\left(E_{gs,0}-E_{\Psi}\right)t}\left|\left\langle \Psi|\Omega\right\rangle \right|^{2}\nonumber \\
 & = & e^{iE_{gs,0}t}  \left\langle \Omega\right|e^{-iHt}\left|\Omega\right\rangle 
\end{eqnarray}
The last expression is the Loschmidt amplitude: the overlap between the state $e^{-iHt}\left|\Omega\right\rangle $, which is the evolution of the initial state under the post-quench Hamiltonian $H$ for time $t$, and the state $e^{-iH_{0}t}\left|\Omega\right\rangle $, which is its evolution under the pre-quench Hamiltonian $H_0$ for the same time $t$. On the other hand, performing the Wick rotation $t\to-iR$, the resulting quantity is the moment generating function of the distribution and can be identified with the partition function of the system confined in a slab of width $R$ with both boundary states equal to $\left|\Omega\right\rangle $. The latter has been extensively studied in the context of the \emph{Casimir effect.} The corresponding free energy per volume $f(R)\equiv-\lim_{L\to\infty}L^{-d}\log G(R)$ can be split into the following three parts according to their behaviour
for large $R$ 
\[
f(R)=f_{b}R+2f_{s}+f_{C}(R)
\]
In the above $f_{b}$ and $f_{s}$ are the bulk and surface contributions,
while the remaining part $f_{C}(R)$ decays for large $R$. In the
original quantum quench problem, $f_{b}$ corresponds to the difference
in the ground state energies of the post- and pre-quench Hamiltonians
per volume, $f_{b}=(E_{gs}-E_{gs,0})L^{-d}$, while $f_{s}$ is related
to the squared norm of the so-called \emph{fidelity} $\left|\left\langle 0|\Omega\right\rangle \right|$
between these states, $f_{s}=-L^{-d}\log\left|\left\langle 0|\Omega\right\rangle \right|$
where $\left|0\right\rangle $ is the post-quench ground state. 
In terms of the probability density function $P(W)$
$f_b$ corresponds to a shift and $f_s$ to the normalization and
the shape is determined only by $f_C$.

The cumulants (equivalent to the moments) of the probability distribution $P(W)$ are given by the logarithmic derivatives of its characteristic function at $t=0$ 
\[
\kappa_{n}=i^{n}\left.\left(\frac{d}{dt}\right)^{n}\log G(t)\right|_{t=0}
\]
Note that since $\log G(t)$ is extensive, all cumulants $\kappa_{n}$
are extensive too, which means in particular, that increasing the
system size $L$, the relative variance  of $P(W)$ tends to zero and in the thermodynamic limit becomes a narrow bell-shaped distribution characterized by its mean value and variance. Note however, that in finite volume a more careful analysis predicts exponential corrections to the cumulants.

The qualitative behaviour of $P(W)$ can be easily inferred by considering
the energy absorption that corresponds to any of the possible transitions
from the initial state to eigenstates of the post-quench Hamiltonian.
The lowest energy absorption corresponds to the transition from the
initial state to the post-quench ground state $\left|0\right\rangle $,
which therefore appears in $P(W)$ as a Dirac $\delta$-function peak
at the value $W=E_{gs}-E_{gs,0}$ with amplitude given by the squared
norm of the fidelity $\left|\left\langle 0|\Omega\right\rangle \right|$
between these states. From now on we will measure the work from the
position of this lowest Dirac $\delta$-function peak, i.e. we will
subtract the ground state energy difference from the work $W$. The
next lowest energy absorption corresponds to the transition to the
lowest post-quench excitation that is allowed. For a massive
post-quench Hamiltonian, like the sinh-Gordon Hamiltonian, the lowest
excitations are separated from the ground state by the energy gap
$m$ i.e. the mass of the lightest quasi-particle excitations. However
since the initial state is translationally invariant and due to the
conservation of the momentum, only transitions to excitations with
zero total momentum are allowed. Furthermore, assuming that both
the pre- and post-quench Hamiltonians are invariant 
under parity transformations of the fields (i.e. if the Hamiltonian $H(\phi)$ 
is invariant under $\phi\to-\phi$ where $\phi$ are the fields describing the system), 
as in our case, there can only be even-particle
excitations in the initial state. This means in particular that there cannot be
any single-particle excitation of momentum zero which would correspond
to a $\delta$-peak at $W$ equal to the particle mass. It is possible
however to have bound-state excitations consisting of even number of
particles with zero total momentum, if the post-quench Hamiltonian contains
such excitations (like in the sine-Gordon model). Apart from such
possible $\delta$-peaks, the lowest allowed transitions 
correspond to the creation of two quasi-particles
with opposite momenta $\pm p$, $p\ge0$. This means that $P(W)$
exhibits a lowest threshold at $2m$ above which there is a continuous
absorption spectrum corresponding to the continuous variable $p$.
The shape of the absorption spectrum above this lowest threshold depends
on the excitation amplitudes, the dimensionality $d$ and the nature
of the excitations (i.e. the statistics, bosonic or fermionic, and
the parameters of their dispersion relation). At the vicinity of the
threshold $P(W)$ exhibits an edge-singularity, typically appearing
as a sharp peak. Provided that the analytic continuation $R\to it$
of the results for the Casimir free energy is valid, it can be shown
that the edge-singularity exhibits universal behaviour controlled
by the large $R$ behaviour of $f_{C}(R)$. In more detail, just above the threshold the distribution exhibits a power-law form 
with an exponent that depends only on the dimensionality and the particle statistics. In 1d the edge exponent is $-1/2$ for bosons and $+1/2$ for fermions (for details, see below). If the excitations are of bosonic nature, there are additional edge-singularity peaks at
positions determined by the thresholds for multi-particle excitation
transitions which overlap with the previous continuous spectrum and
with each other. Obviously there exist peaks for all different types
of particle excitations present in the model, each with its own mass.
Furthermore, any bound-state excitations would manifest themselves
as Dirac $\delta$-function peaks at values equal to their mass (and
any integer multiple of the latter, if they are of bosonic nature), while 
peaks at positions that are odd multiples of the masses are also possible, if odd multi-particle excitations are not excluded by parity symmetry.

In the present case of the sinh-Gordon model and for the particular
choice of initial state, the above general observations apply as follows.
There are no bound-state excitations, therefore there are no $\delta$-function
peaks, other than the ground state one at $W=0$. There is only one
type of excitations with mass $m$, therefore we expect an edge-singularity
above the lowest threshold at $W=2m$. Furthermore, the initial state
consists of pairs of opposite momentum quasiparticle excitations exclusively,
so that there will be additional peaks only at integer multiples of
$2m$. Due to the exponential form, the multi-particle excitations
are all controlled by the single-pair excitation amplitude $K(\theta)$.

\subsection{(Boundary) thermodynamic Bethe ansatz}

Having the definition of the pdf of the work (\ref{defPW}) and the known form of the pre-quench state in terms of post-quench states, i.e. a squeezed state (at least in the important special cases considered in this paper) it seems that obtaining $P(W)$ is just a matter of performing a sum. However, we would like to argue, that it is in fact far from trivial to perform this sum. In fact, in the thermodynamic limit the levels become continuous, the sum turns into an integral and a nontrivial density of states needs to be taken into account in (\ref{defPW}). Furthermore, the squeezed state (\ref{eq:BS}), is not given on a convenient, ordered basis of states,
$$
\vert \theta_1\theta_2\ldots\theta_n\rangle,\qquad \theta_1<\theta_2<\ldots<\theta_n
$$
relative to the post-quench Hamiltonian, therefore to use the definition in addition to the density of states (for all number of particles $n$ possible) we would also need to solve a combinatorial problem to recast the squeezed state in terms of a post-quench basis.

A convenient way to do these tasks is to perform a Fourier transform
and calculate the equivalent Loschmidt amplitude $G(t)$ (\ref{losch}) by analytically continuing the boundary partition function in the slab geometry. This can be obtained in the presence of bulk integrability, i.e. integrable post-quench dynamics, from the thermodynamic Bethe ansatz.

In the remainder of this section we briefly introduce the thermodynamic Bethe ansatz in the presence of boundaries and discuss possible issues when performing analytic continuations.

\subsubsection{Theoretical background}

Consider a $(1+1)$ dimensional quantum system on a finite cylinder of length $R$ and circumference $L$ with
given boundary conditions $|\Omega\rangle$ at the two ends. There are two equivalent quantization
schemes: in the first, time is chosen to run along the axis of the
cylinder (so-called R-channel), while in the second the time direction is perpendicular
to the axis (L-channel). The partition function in the two quantization schemes reads
\begin{align}\label{Zdef}
Z &= \langle \Omega\vert e^{-RH(L)}\vert \Omega\rangle\\
&= \text{Tr}e^{-LH^{\Omega\Omega}(R)}
\end{align}
in effect a projection of the bulk partition function on the state $\vert\Omega\rangle$.
While in the R-channel the boundary conditions can be taken into account as
initial and final states, in the L-channel one must impose instead a finite system size Hamiltonian $H^{\Omega\Omega}(R)$ that depends on the boundary conditions applied at the edges.

Taking now the thermodynamic limit, $L\to\infty$, we get
\begin{align}
Z &= \sum_i |\langle i|\Omega\rangle|^2 e^{-RE_i(L)}\\
&= \sum_j e^{-LE^{\Omega\Omega}_j(R)}
\end{align}
where the first line does not simplify but the second has terms exponentially
more and more suppressed, and it is enough to look at only the first one, i.e.
\begin{equation}
Z\equiv e^{-Lf(R)}\approx e^{-LE_{0}^{\Omega\Omega}(R)}, 
\quad f(R)=f_b R+2f_s+f_C(R),
\end{equation}
where $E_{0}^{\Omega\Omega}(R)$ is the ground state energy in finite volume of a system with boundaries described efficiently by the boundary states $\vert \Omega \rangle$. We remark that the TBA (presented below) yields only the contribution $f_C(R)$ (the bulk term is absent by construction, it can however be extracted through comparison with conformal perturbation theory \cite{KM91} and the surface term corresponds to the normalization of the boundary state $\vert\Omega\rangle$), however this is enough since $f_C(R)$ is the only part that determines the nontrivial shape of $P(W)$.

Now we focus on integrable models, in which the two-body S-matrix, in our case
(theories with a single particle species) a single function $S(\theta)$, characterizes all the dynamics, 
i.e. the higher-body scattering events factorize into two-body collisions. This can be formalized in terms 
of the Faddeev-Zamolodchikov algebra,
\begin{align}
Z(\theta_1)Z(\theta_2)&=S(\theta_1-\theta_2)Z(\theta_2)Z(\theta_1)\\
Z(\theta_1)Z^\dagger(\theta_2)&=S(\theta_2-\theta_1)Z(\theta_2)^\dagger Z(\theta_1)+2\pi\delta(\theta_1-\theta_2)
\end{align}
where the operators $Z^\dagger(\theta)$ generate the space of asymptotic states as
\begin{equation}
\vert\theta_1\ldots\theta_n\rangle=Z^\dagger(\theta_1)\ldots Z^\dagger(\theta_n)\vert0\rangle
\end{equation}

In finite volume $L$, one is subject to the quantization conditions, which can be written in a tractable form, the Bethe-Yang equation
\begin{equation}\label{BetheYang}
e^{imL\sinh\theta_i}\prod_{j\neq i}S(\theta_i-\theta_j)=\pm1,\qquad i=1,\ldots,N
\end{equation}
but only for integrable models. (Note, that in small volume there are exponentially small corrections to these energy levels. The sign corresponds to periodic and anti-periodic boundary conditions, connected to the statistics of particles. In an integrable theory the statistics is reflected in the sign $S(0)=\pm1$. The only known theory with the bosonic sign $S(0)=+1$ is that of free bosons.)
This relation can then be rewritten in terms of particle densities and used as a constraint when performing a saddle point evaluation of the partition function expressed as a functional integral over different configurations. In the thermodynamic limit ($L\to\infty$) the saddle point dominates the functional integral and the resulting expressions for thermodynamic quantities constitute the thermodynamic Bethe ansatz (TBA). The saddle-point solution is equivalent to a superposition state composed of both few- and many-particle excitations, which can explicitly be seen through a multi-particle expansion. Having in mind that the few-particle terms will play an important role in the followings it is important to consider these few-particle terms. At first sight it might seem that the TBA will break down in a limit where such terms dominate or equivalently that these contributions should not be precise in the TBA framework since for the TBA to work the particle density per unit volume must be intensive in $L$, on the other hand to a few-particle state a nonintensive particle density per unit volume corresponds. However, we consider a linear combination of extensively many two-, four-, etc. particle states instead of isolated ones and therefore the corresponding particle density is intensive also in the few-particle limit.

In the presence of boundaries, integrability (in the L-channel) is only preserved if the boundary states respect the very special form of \cite{GZ}
\begin{equation}
\vert
\Omega\rangle=g\exp\left(\int_{0}^{\infty}K(\theta)Z^{\dagger}(-\theta)Z^{\dagger}
(\theta)\right)\vert0\rangle
\end{equation}
and $K(\theta)$ is the analytic continuation of the reflection factor $R$ that describes scattering on the boundary ($K(\theta)=R(i\pi/2-\theta)$). Such boundary integrable states can be incorporated into the TBA construction as a rapidity dependent chemical potential to yield for example\cite{LMSS}
\begin{equation}\label{fCint}
f_C(R) = 
\mp\frac{m}{4\pi}\int_{-\infty}^{\infty}\cosh\theta H(\theta,R)d\theta,
\end{equation}
with
\begin{equation}\label{Hfun}
 H(\theta,R)=\log\left[
1\pm\left|K(\theta)\right|^{2}e^{-\varepsilon(\theta,R)}\right]
\end{equation}
and the pseudoenergy $\varepsilon(\theta,R)$ solves the nonlinear integral
equation
\begin{equation}\label{TBAeq}
\varepsilon(\theta,R)=2mR\cosh\theta+\int_{-\infty}^{\infty}
\Phi(\theta-\theta')H(\theta',R)d\theta' 
\end{equation}
where the logarithmic derivative of the scattering matrix was introduced as
\begin{equation}
\Phi(\theta)=-\frac{1}{2\pi i}\frac{d}{d\theta}\log S(\theta)
\end{equation}

As will become clear later, in terms of the corresponding boundary problem, i.e. in the L-channel, the initial/boundary states relative to quenches cannot preserve integrability (which would correspond to very restrictive choices of the amplitude $K(\theta)$, in particular one that means an infinite energy initial state), therefore it is essential to note, that $K(\theta)$ is in fact arbitrary as long as one's interest is in the quantity (\ref{Zdef}) and is content to remain in the R-channel rather than the boundary problem. This is clear from the derivation of the BTBA which proceeds in the R-channel and then integrability plays a role only in the bulk \cite{LMSS}.

\subsubsection{Existence, uniqueness and analytical properties of the solutions}

The BTBA equation is a non-linear integral equation. On a purely mathematical
basis, it can be shown that for any $R>0$ it possesses a unique real
solution $\varepsilon(\theta,R)$ that is an analytic function of $\theta$
in the neighborhood of the real $\theta$ axis \cite{KM91}. For
$R>0$ the unique solution can be found using the iterative method
which converges uniformly as a function of $R$, thus ensuring that
the solution is analytic also in $R$ in sufficiently small neighborhoods
of any real positive value of $R$. 

Here, we want to analytically continue the BTBA into the complex $R$ plane to obtain the Loschmidt amplitude
\begin{equation}
 G(t)=e^{-Lf_C(it)}
\end{equation}
instead of the partition function. For complex values of $R$ non-analyticities may occur. Using the fact that IQFT can be derived from perturbations of CFTs and in addition the truncation of the Hilbert space of these CFTs, it can be argued that the possible non-analyticities of $\varepsilon(\theta,R)$ as a function of $R$ are in general square-root branch points that appear whenever two eigenvalues of the perturbing operator become degenerate \cite{KM91}. In the thermodynamic limit such square-root branch points accumulate around the critical point of the theory as explained in the Yang-Lee theory of phase transitions. It is interesting to note, that in the corresponding lattice model the branch points become singularities, giving rise to the Fisher zeroes of the partition function \cite{KM91}.

We point out that in the presence of boundaries the analytic structure can and does change. In fact, for the sinh-Gordon model there are branch points on the imaginary axis for the TBA without boundaries, while when we used the boundary state relative to a quench, we no longer found any branch points. One can begin to understand this in terms of the corresponding Fisher zeroes of the lattice partition function: if one does a quench originating from one phase and arriving in a different phase, one may expect (not always) a dynamical phase transition in the time evolution, which is governed by the Fisher zeroes \cite{HPK}. With a reversed logic, when quenching inside the same phase, as in our case, a dynamical phase transition is not expected (although in some cases they were observed also without crossing a phase transition line, see \cite{AS, vd-14}), thus neither are Fisher zeroes, i.e. branch points.

Moving on now to the solution of the BTBA for complex $R$, it is a priori obvious, that in the left half-plane the BTBA equations cannot be solved by means of the usual iterative scheme with the free solution as initial step, since in this case the integral equation would diverge. Therefore existence and uniqueness of the solution is not guaranteed, at least not based on the standard iterative approach. Luckily, this is not needed for our purposes, since the analytical continuation is all done inside the right half-plane. In the latter the iterative method is valid and one only needs to track possible singularities that may block the analytical continuation. Viewing these singularities in the complex rapidity plane, where they appear as zeroes of the logarithm in (\ref{TBAeq}), we see that when we vary the value of $R$, one such singularity may approach the contour of integration from one side. Even when it crosses the real rapidity axis though, we can deform the integration contour off the axis on the 
opposite side, where the integrand is analytic, therefore keeping the solution of the TBA equation analytic in $R$. If we insist to write the TBA equation with the integration contour along the real axis, we have to modify it by including the contribution of the logarithmic singularity, in order to stay analytically connected to the real-$R$ TBA equation. If however, while varying $R$, a pair of such singularities approach the integration contour at the same point from both sides, such a contour deformation is not possible and the TBA equation exhibits non-analytic behaviour for that value of $R$. These are called \emph{pinching singularities}. This is a problem familiar from the study of the excited state energies \cite{DT1,DT2,DT3}.

In the simple case of the sinh-Gordon with the quench boundary state, such subtleties do not arise. Approaching the imaginary axis is still nontrivial: even if the boundaries regularize an iteration approach, one is still left with oscillating integrals instead of exponentially decaying ones. This feature poses a serious numerical bottleneck, especially for $R$ with larger absolute values. We are also interested in the work statistics, therefore knowledge of the Loschmidt amplitude on a large domain is necessary. In section 3.2 we propose an evaluation technique that we applied successfully to solve the BTBA for $R=it$, with $t$'s being effectively arbitrarily large.

\subsection{The initial state}

In this section we discuss the role of the initial state in the calculation of the work statistics in quantum quenches. We first summarize results for free models, bosonic and fermionic, that will help us better understand the results for integrable models in general and the shG model in particular.

\subsubsection{Free theories}

For mass quenches the boundary state in both the bosonic and fermionic theories can be written as
\begin{equation}
\vert \Omega\rangle =
N\exp\left(\int_0^\infty \frac{dp}{2\pi} K(p) A_{-p}^{\dagger}A_{p}^{\dagger}
\right)\vert0\rangle
\end{equation}
where $A_{p}^{\dagger}$ is the free boson/fermion creation operator in momentum representation. For free bosons the amplitude $K_{FB}(p)$ is  \cite{SFM,SGS}
\begin{equation}
K_{FB}(p)= -\frac{E_{0p}-E_{p}}{E_{0p}+E_{p}}
\end{equation}
where $E_{0p} = \sqrt{p^{2}+m_{0}^{2}}$ and $E_{p} = \sqrt{p^{2}+m^{2}}$. Substituting $p = m\sinh\theta$, the above equation can be written in terms of rapidities as
\begin{align}
K_{FB}(\theta)&= -\frac{\sqrt{\sinh^2\theta+ (m_0/m)^2}-\cosh\theta}{\sqrt{\sinh^2\theta+ (m_0/m)^2}+\cosh\theta} \\
&= -\frac{\sinh(\theta - \varphi(\theta))}{\sinh(\theta + \varphi(\theta))} \label{eq:KFB}
\end{align}
where 
\begin{equation}
\sinh\varphi(\theta) \equiv \frac{m}{m_0} \sinh\theta
\end{equation}

For free (Majorana) fermions instead the boundary state amplitude is \cite{SE,SFM}
\begin{equation}
K_{FF}(p) = i \frac{\sqrt{(E_{0p} - p) (E_{p} +p)} - \sqrt{(E_{0p} + p) (E_{p} - p)}}
{\sqrt{(E_{0p} + p) (E_{p} + p)} + \sqrt{(E_{0p} - p) (E_{p} - p)}}
\end{equation}
or in terms of rapidities
\begin{equation}
K_{FF}(\theta)= i\frac{\sinh\left(\frac{\theta-\varphi(\theta)}{2}\right)}
{\cosh\left(\frac{\theta+\varphi(\theta)}{2}\right)}
\end{equation}

\subsubsection{Initial state after a quantum quench in the shG model}

The sinh-Gordon model is a simple model from the TBA point of view,
however it contains genuine, strong interaction and is an interesting testing
ground for our ideas. The Lagrangian reads
\begin{equation}
\mathcal{L}=\frac{1}{4\pi}(\partial_\nu\phi)^2+2\mu\cosh(2b\phi)
\end{equation}
and describes an integrable field theory with a single particle species with physical mass
\begin{equation}
 m=4\sqrt{\frac{\mu\sin B\pi}{1-B}},\quad B=\frac{b^2}{b^2+1}
\end{equation}
and scattering amplitude and phase shift
\begin{align}
S(\theta)&=\frac{\sinh\theta-i\sin \pi B }{\sinh\theta+i\sin \pi
B },\\
\Phi(\theta)&=-\frac{1}{\pi}\frac{\sin(B\pi)\cosh\theta}{
\sin^2(B\pi)+\sinh^2\theta}
\end{align}
The Fourier transform of the phase shift reads
\begin{equation}
\tilde{\Phi}(t)=-\frac{\cosh\frac{a\pi t}{2}}{\cosh\frac{\pi t}{2}},\qquad
a=1-2B
\end{equation}
Notice the weak/strong coupling duality $B\leftrightarrow 1-B$.

In general, determining the initial state $|\Omega\rangle$ after a quantum quench in an IQFT, i.e. calculating the amplitudes of excitations (in the post-quench basis) contained in the initial state, is a difficult problem. One has to extract this information from the defining property of the initial state that it is annihilated by the annihilation operators of the pre-quench Hamiltonian, whose expression in terms of the post-quench creation and annihilation operators is generally unknown \cite{SFM}. However, in the case where the pre-quench Hamiltonian is non-interacting ($g_0=0$) the above requirement reduces to the simpler condition
\begin{equation}
\left(\phi(p)+\frac{1}{\sqrt{p^2 + m_0^2}}\Big[\phi(p),H\Big]\right) |\Omega \rangle = 0
\end{equation}
valid for all momenta $p=m\sinh\theta$. In the latter, $m_0$ is the pre-quench mass and $\phi(p)$ the Fourier transform of the physical field $\phi(x)$. If we formally expand $|\Omega\rangle$ as
\begin{multline}
|\Omega \rangle =  |\text{vac}\rangle + K_1(0)|0\rangle + \int_0^\infty d\theta K_2(\theta)|{-\theta},\theta\rangle \\
+ \sum_{n=3}^\infty \int d\theta_1... d\theta_n\; K_n(\theta_1,...,\theta_n) |\theta_1,...,\theta_n\rangle 
\end{multline}
(up to an overall normalization factor) then, applying suitable test states on the left of the former equation, we can derive equations that must be satisfied by the amplitudes $K_n$ of the excitations in $|\Omega\rangle$ \cite{STM}. These equations are integral equations consisting of an infinite series that involves amplitudes $K_n$ of all orders, as well as the form-factors of the field $\phi$. Obviously it is impossible to solve these equations, unless we have some good ansatz for the solution. Based on general properties of the pre-quench ground state (translational and parity invariance) it can be shown that all odd amplitudes $K_{2n+1}$ vanish. Furthermore, as mentioned in the introduction, we know that the solution in the limit of infinite $m_0$ is the Dirichlet state defined by $\phi|D\rangle =0$ whose exact form is already known by other means \cite{GZ,G,CT}: it is of the exponential form (\ref{eq:BS}) with amplitude $K_D(\theta)$ given by
\begin{equation}
K_D(\theta)=i\tanh(\theta/2)\left(\frac{1+\cot(\pi B/4-i\theta/2)}{1-\tan(\pi B/4+i\theta/2)}\right) \label{eq:KD}
\end{equation}
It can be verified that the amplitudes of $|D\rangle$ satisfy the integral equations as expected. However this state exhibits ultraviolet divergences in the calculation of observables, due to the non-decaying behaviour of $K_D(\theta)$ for large $\theta$ or, equivalently, due to the fact that the natural ultraviolet bound of excitations is of the order of $m_0$ which is considered to be infinite. In particular, in the calculation of the Loschmidt echo and work statistics using the BTBA, a sufficiently fast ultraviolet decay of the amplitude $K$ is necessary in order to ensure the convergence of the integrals in (\ref{fCint}) and (\ref{TBAeq}). Indeed, while for real $r$ the integrand decays exponentially even for $K$ that does not decay for large $\theta$, for imaginary $r$ instead it exhibits oscillatory behaviour unless $K$ decays itself. Assuming a large but not infinite $m_0$ modifies the form of the state and, from the integral equations, it can be shown \cite{STM} that to a good approximation the 
modified state can be described by the same exponential form (\ref{eq:BS}) with a new amplitude 
\begin{equation}
K(\theta)= -K_D(\theta) K_{FB}(\theta) \label{eq:Knew}
\end{equation}
Note that the large $\theta$ behaviour of this amplitude is the same as that of $K_{FB}$ i.e. it decays as $e^{-2\theta}$. This solves the problem of ultraviolet divergences of the Dirichlet state. In our BTBA calculations we will use exactly this choice of initial state. The above arguments can be generalized to other relativistic IQFTs.

\section{Results}
\subsection{The edge singularity of the work statistics in IQFT}

Before we proceed to the numerical analysis of the shG model, we report a general exact result for the shape of the lowest peak of the work probability distribution $P(W)$ in any IQFT in large $L$: the part $2m<W<4m$ (for a single particle theory) is entirely determined by $K(\theta)$ since in this region there can only be two particles in the system with opposite momenta, which never scatter off each other thus the form of the scattering matrix is irrelevant. 

\subsubsection{BTBA analysis}

In this section using the BTBA we obtain $P(W)$ analytically in the region $2m<W<4m$ and provide a systematic method to calculate it in terms of multiple integrals for $W>4m$. This calculation relies on that ultimately we only need the Fourier transform of the solution of the BTBA with respect to $t$, and that each step of an iterative solution provides exactly the Fourier components in sequential frequency windows. This property is true because of the gap in the excitation spectrum. Solving the BTBA equation exactly in energy space, i.e. obtaining the Fourier transform of the solution with respect to $t$, is also interesting on its own right, since incidentally we can provide a systematic method to obtain the solution of a nonlinear integral equation exactly, although in terms of multiple integrals.

First of all it is useful to introduce the difference between the pseudoenergy $\varepsilon(\theta,it)$ for the interacting and non-interacting scenarios, i.e. we define
$$
y(\theta,t)=\varepsilon(\theta,it)-2imt\cosh\theta
$$
This function satisfies the nonlinear integral equation
\begin{multline}
y(\theta,t)=\int d\theta'\Phi(\theta-\theta')\\
\times\log\left(1+|K(\theta')|^2e^{-2imt\cosh\theta'}e^{-y(\theta',t)}\right)
\end{multline}
equivalent to (\ref{TBAeq}). Now we expand the logarithm and the exponential $e^{-y}$ to obtain
\begin{multline}
y(\theta,t)=\sum_{n=1}^\infty\sum_{\ell=0}^\infty\frac{(-n)^\ell}{n \ell!}\int d\theta'\Phi(\theta-\theta')|K(\theta')|^{2n}\\
\times e^{-2inmt\cosh\theta}y(\theta',t)^{\ell}
\end{multline}
and we begin an iteration $y_{(n+1)}=y_{(n+1)}[y_{(n)}]$ with initial approximation $y_{(0)}\equiv0$. 

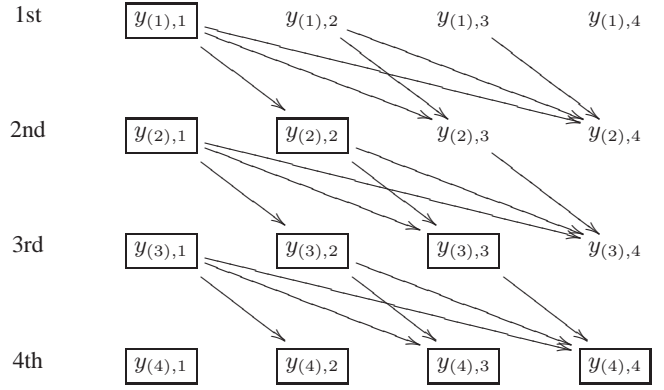
\begin{figure}
\begin{displaymath}
    \xymatrix{
        \text{1st} & \boxed{y_{(1),1}} \ar[dr] \ar[drr] \ar[drrr] & y_{(1),2} \ar[dr] \ar[drr]& y_{(1),3} \ar[dr]& y_{(1),4} \\
        \text{2nd} &  \boxed{y_{(2),1}} \ar[dr] \ar[drr] \ar[drrr] & \boxed{y_{(2),2}} \ar[dr] \ar[drr]& y_{(2),3} \ar[dr]& y_{(2),4} \\
        \text{3rd} &  \boxed{y_{(3),1}} \ar[dr] \ar[drr] \ar[drrr] & \boxed{y_{(3),2}} \ar[dr] \ar[drr]& \boxed{y_{(3),3}} \ar[dr]& y_{(3),4} \\
        \text{4th} & \boxed{y_{(4),1}}& \boxed{y_{(4),2}} & \boxed{y_{(4),3}}& \boxed{y_{(4),4}}
        }
\end{displaymath}
\caption{First four iterations of the solution of the BTBA equation showing the terms in the partitioning (\ref{parti}) becoming successively exact (highlighted). $y_{(i),n}(\theta,t)$ is the n-th term in the partitioning after the i-th iteration. The arrows show which terms in the iteration influence which. The first term is already exact after the first step and the others become exact if all the terms influencing them have already become exact by the previous step in the iteration.}\label{figTBA}
\end{figure}

Since $\cosh\theta\geq1$ for all $\theta$, if we look at the frequency components of the solution, we see that they become successively exact in the windows $\omega\in[2m,4m]$, $[4m,6m]$, $\ldots$ after the first, second, etc. iterations. Also, we see that the frequencies $\omega<2m$ do not contribute at all. We can formalize this by considering a partition of the solution
\begin{equation}\label{parti}
y(\theta,t)=\sum_{n=1}^\infty y_n(\theta,t)
\end{equation}
so that the Fourier transform $\bar{y}_n(\theta,\omega)$ of the term $y_n(\theta,t)$ with respect to $t$ disappears in the window $\omega<2nm$ and is non-zero for $\omega>2nm$, i.e.
$$
\bar{y}_n(\theta,\omega)\sim \theta(\omega-2nm)
$$
In Fig. \ref{figTBA} we show how the successive terms in the partitioning (\ref{parti}) become successively exact.

We write here the exact solution in the first two windows of $W$, but in principle they can be generated in terms of multiple integrals for arbitrary windows,
\begin{align}
 y_1(\theta,t)&=\int d\theta'\Phi(\theta,\theta')|K(\theta')|^2e^{-2imt\cosh\theta'}
\\
 y_2(\theta,t)&=\frac{1}{2}\int d\theta'\Phi(\theta,\theta')|K(\theta')|^4e^{-4imt\cosh\theta'}\nonumber\\
 &\qquad-\int d\theta'\Phi(\theta,\theta')|K(\theta')|^2e^{-2imt\cosh\theta'}\nonumber\\
 &\qquad\quad\times\int d\theta''\Phi(\theta',\theta'')|K(\theta'')|^2e^{-2imt\cosh\theta''}
\end{align}

With the solution of the BTBA at hand we calculate the work pdf as
\begin{align}
&P(W)=e^{-2f_s}\int \frac{dt}{2\pi}e^{iWt-Lf_C(it)}\\
&\,\,=e^{-2f_s}\left[\delta(W)-L\bar{f}_C(W)+\frac{L^2}{2}(\bar{f}_C\ast\bar{f}_C)(W)+\ldots\right]
\end{align}
by expanding the exponential. Since the solution of the BTBA admits the partitioning (\ref{parti}) both $\bar{f}_C$ and $P(W)$ have a corresponding form, i.e.
\begin{align}
\bar{f}_C(W)&=\sum_{n=1}^\infty \bar{f}_{C,n}(W)\theta(W-2nm)\\
P(W)&=e^{-2f_s}\left[\delta(W)+\sum_{n=1}^\infty p_{n}(W)\theta(W-2nm)\right]
\end{align}
In the latter we can identify that each term $p_n$ corresponds to the production of $2n$ particles. In particular, the 2-particle production contribution reads
\begin{equation}
 p_1(W)=\frac{mL}{4\pi}\int d\theta\cosh(\theta)|K(\theta)|^2\delta(W-2m\cosh\theta)
\end{equation}
and the four-particle
\begin{multline}
 p_2(W)=\frac{(mL)^2}{2(4\pi)^2}\int d\theta\int d\theta'\cosh(\theta)\cosh(\theta')\\
 \times|K(\theta)|^2|K(\theta')|^2
 \left[1+\frac{4\pi\delta(\theta-\theta')}{mL\cosh\theta'}-\frac{8\pi\Phi(\theta-\theta')}{mL\cosh\theta'}\right]\\
 \times\delta(W-2m\cosh\theta-2m\cosh\theta')
\end{multline}
where the first term comes from the self-convolution of the lowest term in $f_C$, while the second two originate from higher corrections to $f_C$, the final in particular comes from $y_1(\theta,t)$. We can see now, that although we obtained $P(W)$ analytically, or at least we know how to obtain it for arbitrary $W$, a numerical approach is still useful, since the actual evaluation of higher terms becomes increasingly cumbersome.

Nevertheless, the interesting low-energy part of $P(W)$ can be obtained extremely simply and analytically, the result reads
\begin{equation}\label{edge}
 P(W<4m)=\frac{e^{-2f_s}LW}{8\pi\sqrt{W^2-4m^2}}|K(\text{arcosh}{\textstyle\frac{W}{2m}})|^2
\end{equation}
This means in particular, that the first edge is explicitly independent of the $S$-matrix and the interaction plays a role only through the expansion of the initial state in terms of post-quench degrees of freedom.

\begin{figure}[h]
  \centering
  \includegraphics[width=0.45\textwidth]{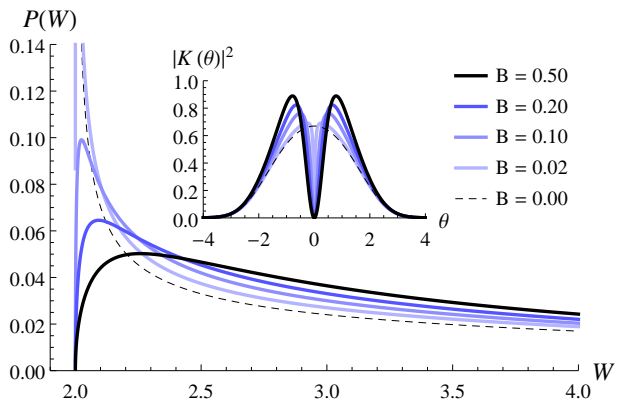}
  \caption{(Color online) Edge singularity and low energy behavior of the probability distribution function calculated analytically for the sinh-Gordon model with different interaction strengths at $L=5m^{-1}$ and corresponding to the initial/final mass ratio $m_0=10m$. (Energy is measured in units of the elementary mass $m$.)  (Inset) Particular forms of the function $|K(\theta)|^2$ describing the initial state.}\label{firstedgefig}
\end{figure}

In Fig. \ref{firstedgefig} we show the resulting analytic plots of the first edge in the sinh-Gordon model using various interaction strengths. One can check, that the edge exponent becomes exactly $+1/2$ contrary to the free boson case, where it is $-1/2$.
This is a direct consequence of the behavior of $|K(\theta)|^2$ (also depicted) near the origin: while in the free case it is a Gaussian and finite at $\theta=0$, when the interaction is turned on it becomes zero for $\theta=0$ forbidding the creation of identical particles.

\subsubsection{Direct analysis via the Bethe-Yang equation}

Before proceeding to the numerical calculation that is capable to access the work pdf for arbitrary energy, we demonstrate that the result (\ref{edge}) for the low energy part of $P(W)$ can be reproduced using its definition and the Bethe-Yang equation. In fact, since the amplitudes $\vert\langle\Psi\vert\Omega\rangle\vert$ are accessible by the form factor approach of \cite{STM} all we need is the distribution of the post-quench states, which can be inferred from the Bethe-Yang equation. Moreover, the form factor approach is capable to provide the amplitudes $K_n(\theta_1,\ldots,\theta_n)$ also in the more general case of arbitrary initial model parameters.

We begin with the definition, which takes the form
\begin{equation}
P(W<4m)=e^{-2f_s}\sum_{\theta_i} \delta(W-E_{\theta_i})\underbrace{\vert\langle\theta_i,-\theta_i\vert\Omega\rangle\vert^2}_{|K_2(\theta_i)|^2}
\end{equation}
We used, that states with energy smaller than $4m$ are all two particle states and that the initial state $\vert\Omega\rangle$ is translationally invariant. In finite volume the admissible set of $\theta_i$ is given by solutions of the Bethe-Yang equation (\ref{BetheYang}), in this case of the form
\begin{equation}
 m\sinh\theta_n-\frac{i}{L}\log S(2\theta_n)=\frac{2\pi n}{L},\qquad n\in\mathbb{Z}
\end{equation}
Since $\log S(2\theta)$ is bounded, in large volume the expression simplifies to 
\begin{equation}
 m\sinh\theta_n=\frac{2\pi n}{L},\qquad n\in\mathbb{Z}
\end{equation}
and the levels can be substituted with a continuous distribution of states with density
\begin{equation}
 \rho(\theta)=\frac{mL}{2\pi}\cosh\theta
\end{equation}
and the sum can be rewritten as an integral,
\begin{equation}
P(W<4m)=e^{-2f_s}\int_0^\infty d\theta\rho(\theta) \vert K_2(\theta)\vert^2 \delta(W-2m\cosh\theta)
\end{equation}
giving rise to the same expression (\ref{edge}) with the two-particle amplitude $K_2(\theta)$.

We stress here, that the formula (\ref{edge}) remains valid for arbitrary initial mass and interaction strength, in which case $K_2(\theta)$ does not follow the regularized Dirichlet form, but it needs to be calculated by solving the integral equations of \cite{STM}. We also note, that in general, (\ref{edge}) should remain valid as long as the particle picture is valid, i.e. even in the presence of a small integrability breaking perturbation as long as the energy spectrum has the same structure, e.g. production of new particles is not induced at these energies.

To proceed further in the pdf, to higher values of $W>4m$, such a simple analysis is no longer possible. Already for the four-particle states there is a much more prominent influence of the scattering matrix on the values of allowed rapidities in the Bethe-Yang approach and inferring the density of states is not trivial anymore. Therefore, it is indeed necessary to find an alternative approach (here the BTBA) for the work pdf other than direct evaluation of the definition.

\subsection{Numerical results}

\subsubsection{Free theories}

For free bosonic theories the statistics of the work done was already studied earlier \cite{SGS}. It is however useful to present the results in terms of the TBA. The free (i.e. $\Phi(\theta)=0$) BTBA prescribes $\varepsilon(\theta,Rm)=2mR\cosh\theta$ and 
\begin{multline}
f_{C}(R) = \\
\pm\frac{m}{4\pi}\int_{-\infty}^{\infty}\cosh\theta\log\left[1\mp\left|K(\theta)\right|^{2}e^{-2mR\cosh\theta}\right]d\theta
\end{multline}
where the upper and lower signs correspond to bosons and fermions, respectively, and 
account for $S(0)=\pm1$. The only difference, apart from the signs, is in the form of the boundary state describing the particular physical process, i.e. the quench.

\begin{figure}[ht]
  \centering
  \includegraphics[width=0.45\textwidth]{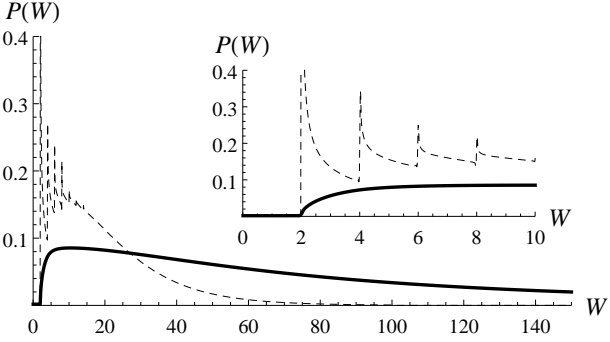}
  \caption{(Color online) Probability density functions $P(W)$ at $L=5m^{-1}$ for the free boson (dashed line) and free fermion (full line) theories for a mass quench with $m_0=10m$. The latter density function was multiplied by a factor of 2 for better visibility. (Energy is measured in units of the elementary mass $m$.)
  (Inset) Blow-up of the initial part of $P(W)$.}\label{fig1}
\end{figure}

In figure \ref{fig1} we show the work statistics for a mass quench in the free theories. The two distributions are very different: the tails of the distributions are decaying differently and for low energies the bosonic function contains a characteristic sequence of edge singularities, which are absent from the fermionic distribution. The differences can be understood by studying the initial states. The corresponding $K$ functions are markedly different from each other: the fermionic is  zero at $\theta=0$, that signals the fermionic nature of the scattering (multiple particle excitations of equal rapidities are suppressed). It also has a different tail, namely while $K$ decays as $e^{-2\theta}$ for bosons, it only goes as $e^{-\theta}$ for fermions. This causes the distribution function to be more extended, having a more slowly decaying tail. It is also easy to check that the edge exponent is $-1/2$ for bosons and $+1/2$ for fermions. These differences can all be seen in figure \ref{fig1}.

We note, that the evaluation of $f_C(R)$ for imaginary $R$ is not completely trivial as the integral is highly oscillatory. We noticed, however, that the integration contour can be Wick rotated. The most useful form is
\begin{multline}
f_C(it)=\pm2im\int_{-\infty}^{0}\frac{1+iy}{\sqrt{2iy-y^{2}}}\\
\times\log\left[
1\mp\left|K(\text{\text{arcosh}}(1+iy))\right|^{2}e^{-2imt(1+iy)}\right]dy
\end{multline}
valid for $t>0$. For $t<0$ on can use the reflection principle $f_C(-it)=f_C(it)^{*}$.

\subsubsection{Numerical solution of BTBA}

In our numerical calculations we use the important fact, that the BTBA integral
equation is a convolution equation in the variable $\theta$, therefore it is 
simpler and practically easier to solve it in Fourier space. In Fourier 
space with respect to the first variable $\theta$ the BTBA equation reads
\begin{equation}
\tilde y(\tau,t)=\tilde\Phi(\tau)\tilde H(\tau,it)
\end{equation}
We solve this by simply iterating the equation (as $\tilde y_{(n+1)}=\tilde\Phi\tilde H_{(n)}$ with $y_{(0)}\equiv0$) until we arrive sufficiently near to a fixed point in the function space (the target was to reach 1 \% change in $L^1$ norm). The iterative step is performed in Fourier space, whereas for the determination of $\tilde{H}_{(n)}$ from $\tilde{y}_{(n)}$ we have to go back to the $\theta$ variable. Therefore we discretized the problem and performed the Fourier transforms by FFT. The precision of the solution can be checked by substituting the resulting $y(\theta,R)$ function to the original integral equation. Precision is controlled by the cut-offs in $\theta$ and $\tau$ spaces beyond all the functions are supposed to be zero, the discretization steps, and the tolerance target.

From the analysis of Section 3.1.1 we expect that for increasing $t$ less and less iterations are neccessary to obtain a precise solution of the BTBA equation, which property we in fact observed when doing the actual numerics. This is because the large $t$ behavior of $y(\theta,t)$ is controlled by the small $\omega$ behavior of the Fourier transform with respect to $t$, $\bar{y}(\theta,\omega)$, which we shown to become exact after already one iteration (or for larger $\omega$ (relevant for smaller $t$) a few, but finite iterations).

There is a problematic point that was already mentioned: while for real temperatures the BTBA and the integral for the ground state energy involve exponentially suppressed kernels, for imaginary temperatures these turn into oscillating ones, which however still get suppressed for large $\theta$. When iterating the BTBA in Fourier space this does not pose a serious problem, however when calculating the ground state energy from the pseudoenergy it does. This probably stems from that in the former case we have to Fourier transform a highly oscillatory function, while in the latter we rather have to Laplace transform it. To overcome this difficulty we analyzed the BTBA and determined the following, previously seemingly unnoticed property.

Consider a function $f(\theta)$ and its Fourier transform $\tilde f(\tau)$, so that
\begin{equation}
\tilde f(\tau)=\int_{-\infty}^\infty e^{i\tau\theta}f(\theta)d\theta,\qquad f(\theta)=\frac{1}{2\pi}\int_{-\infty}^\infty e^{-i\tau\theta}\tilde f(\tau)d\tau
\end{equation}
It is well known, but also easy to see from the definition that exponential decay, i.e. $f(\theta)=f_1 e^{-b_1\theta}+\ldots$ as $\theta\to\infty$ corresponds to a pole of $\tilde f(\tau)$ located at $\tau=-ib_1$ with residue $2\pi f_1$, which is also the closest one to the real line. (Corrections to the first term of the asymptotic formula correspond to other poles, further away from the real line.) In our cases, in particular,
\begin{eqnarray}
\Phi(\theta)=\Phi_1 e^{-\theta}+\ldots,\qquad \theta\to\infty\\
H(\theta,R)=H_1(R) e^{-2\theta}+\ldots,\qquad \theta\to\infty
\end{eqnarray}
where the latter comes from the asymptotics of the specific form of $K$ relative to quenches. Since $\tilde y=\tilde\Phi\tilde H$, $\tilde y$'s closest pole to the real line lies at $\tau=-i$ (determined by $\Phi$) and thus
\begin{equation}
y(\theta,t)=\Phi_1 e^{-\theta} \tilde H(-i, it)+\ldots,\qquad \theta\to\infty
\end{equation}
conversely
\begin{equation}
\tilde H(-i,it)=\lim_{\theta\to\infty}\frac{y(\theta,t)}{\Phi(\theta)}
\end{equation}
Now, using the definition of $\tilde H$,
\begin{equation}
\tilde H(-i,it)=\int_{-\infty}^\infty e^{-\theta}H(\theta,it)d\theta=\int_{-\infty}^\infty \cosh\theta H(\theta,it)d\theta
\end{equation}
where the second equality if due to $H(\theta,it)$ being even in $\theta$, we find that
\begin{equation}\label{asyeval}
f_C(R)=-\frac{m}{4\pi}\lim_{\theta\to\infty}\frac{y(\theta,-iR)}{\Phi(\theta)}
\end{equation}
That is, we have now an evaluation protocol for $f_C(R)$ by obtaining it directly from the asymptotics of $y(\theta,-iR)$ without the need for a further integral transformation.

\subsubsection{Loschmidt echo and work statistics in the shG model}

We now focus on the specific case of a quantum quench in the shG from zero initial coupling and large initial mass to any coupling and mass, as introduced before. We perform numerical calculations of the Loschmidt echo as a function of time, the partition function $Z(z)$ given by (\ref{eq:Z}) in the complex $z$-plane, the pdf of the work done and the dependence of the mean work on the post-quench coupling and mass.

\begin{figure}[ht]
\centering
 \includegraphics[width=0.48\textwidth]{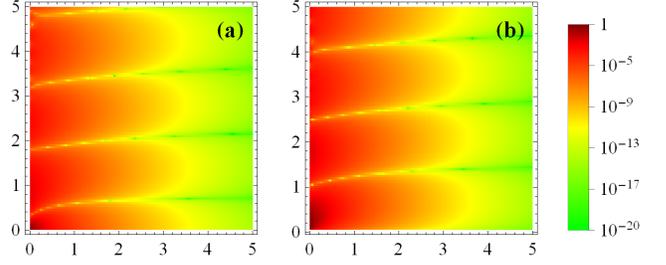}
 \caption{(Color online) Real (a) and imaginary (b) parts of $f_C(z)$ in the upper right quarter of the 
complex plane at the particular choices of the parameters $m_0=10m$, $B=0.5$. The lower
right quarter is obtained by the reflection principle $f_C(z^*)=f_C(z)^*$.}\label{plane}
\end{figure}

To get the Loschmidt echo
$$
L(t)=|G(t)|^2=e^{-L(f(it)+f(-it))}=e^{-2L\text{Re}\,f(it)}
$$
we continue analytically the partition function $Z(z)$ from the positive real line $z\in\mathbb{R}^+$, where the TBA equation is known to yield a unique result, to the imaginary line $z\in i\mathbb{R}$. As we discussed before it is possible that for some lines in the $z$-plane $Z(z)$ develops a branch cut corresponding to logarithmic divergences in the function $H(\theta,z)$. 
For some values of $t\in\mathbb{R}$ we performed a numerical continuation of $f_C(t)$ to $f_C(it)$ following lines $z=t e^{i\phi}$ along $\phi\in[0,\pi/2]$ and looked at every step for the zeroes of the argument of the logarithmic term $H(\theta,z)$. We did not find any zeroes crossing the integration contour. Therefore we elected to simply solve the original BTBA equation iteratively with initial condition $y_{(0)}(\theta,-iz)\equiv0$ for a dense grid in the upper right quarter of the complex plane. Since the resulting function (depicted in Fig. \ref{plane}) is smooth and satisfies the Cauchy-Riemann equations we concluded that it is in fact the analytic continuation of $f_C(z\in\mathbb{R})$. By this argument we can be sure that in our case the solution on the imaginary line obtained from the original BTBA equation with initial condition $y_{(0)}(\theta,t)\equiv0$ is the true Loschmidt amplitude.

\begin{figure}[ht]
\centering
  \includegraphics[width=0.45\textwidth]{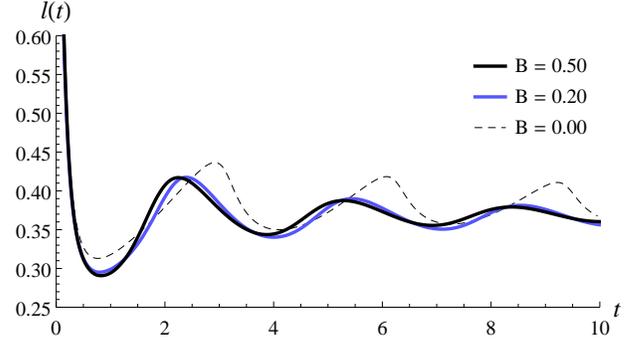}
 \caption{(Color online) Loschmidt echo per unit volume $l(t)=L(t)^{1/L}$ in the sinh-Gordon model after a quench from $m_0=10m$ to $m$ and $B=0$, $0.2$, $0.5$.  (Time is measured in units of the elementary mass $1/m$.)}\label{losch}
\end{figure}

The Loschmidt echo $L(t)$ is depicted in figure \ref{losch} for different post-quench model parameters.

\begin{figure}[h]
  \centering
  \includegraphics[width=0.45\textwidth]{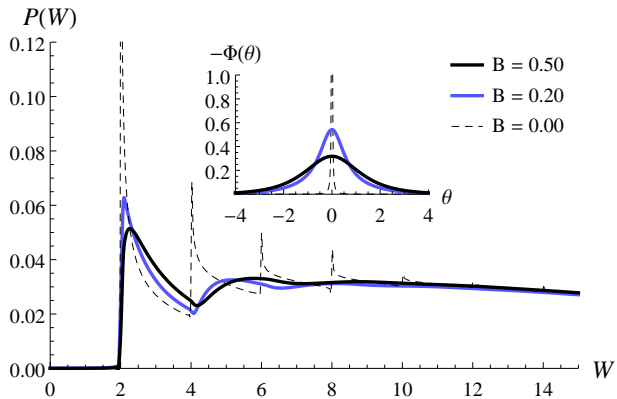}
  \caption{(Color online) Probability density functions at $L=5m^{-1}$ and different interaction strengths with the same initial/final mass ratio $m_0=10m$. (Inset) Particular forms of the function $-\Phi(\theta)$ for these parameter settings. (Energy is measured in units of the elementary mass $m$.)}\label{PW}
\end{figure}

\begin{figure}[h]
\centering
 \includegraphics[width=0.45\textwidth]{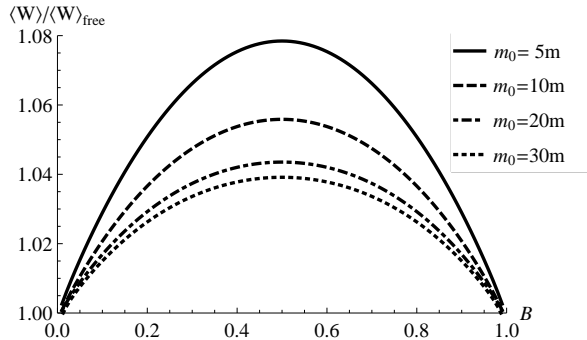}
 \caption{Mean work as a function of the coupling constant compared to the free
boson result at different initial/final mass ratios. Note that the volume cancels out
from the ratio because of extensivity.} \label{mean}
\end{figure}

With an accurate Loschmidt amplitude calculated on a domain of $t\in[0,100]$ we are now in a position to proceed to the calculation of the work pdf. Results for different post-quench parameters are shown in figure \ref{PW}. We see, that while the statistics as a whole departs only slightly from the free boson result (see also figure \ref{mean}, where the departure of the mean work is depicted relative to the free result), the details change qualitatively. The sequence of sharp edge singularities  with negative exponents at the openings of new channels, characteristic to the free bosons, become edges with positive exponents, and also less and less pronounced as the interaction strength is increased. Near at first edge $K(\theta\to0)$ matters the most and we find a behavior more similar to that of free fermions. The edge exponent becomes exactly the fermionic value $+1/2$ instead of $-1/2$, characteristic to free bosons. On the other hand, while for the free fermions there is just one pronounced edge at $2m$,
consistent with the Pauli principle\footnote{At $W=4m$ a new channel opens, that involves 4 particles being created after the quench. This channel is however completely suppressed for fermions at $W=4m$ because it would mean the creation of 4 identical, stationary fermions.}, for the sinh-Gordon bosons there are still multiple peaks. In the low interaction strength limit, the results tend to the free boson case, and in fact one can easily see using $\lim_{B\to0}\Phi(\theta)=-2\pi\delta(\theta)$ in equations (\ref{fCint}), (\ref{Hfun}), (\ref{TBAeq}), that $f_C(r)$ goes to the free boson result.

\section{Conclusions}

In this paper we studied the statistics of the work done when a quantum quench is performed on an integrable field theory and the Loschmidt echo after the quench. Until now the work statistics was only obtained for non-interacting models or those that can be mapped to such. In the present paper one of our main objectives was to see the effect of turning on a genuine interaction.

Using the equivalence of the work probability density function (pdf) and its characteristic function, the Loschmidt amplitude, we proposed to find the pdf from the analytic continuation of the partition function provided by the boundary thermodynamic Bethe ansatz (BTBA). Our motivation was given partially by recent advances on the expansion of the initial quench state on the post-quench basis, in particular that for certain kinds of quenches the initial state is similar to boundary integrable states, for which the BTBA can be used.

We proposed that because of integrability in these models the initial part of the pdf is universal, in the sense that it only depends on the initial state and not the particulars of the scattering. That is, if two different systems are prepared in the same initial state (in terms of asymptotic particles), the initial part of the pdf of the required work is the same. For the case of one particle species we gave a formula  that gives the first peak of the pdf analytically.

Then we proceeded to calculate the Loschmidt echo and the whole pdf for a particular model. The difficulty of such a calculation stems form the fact that an accurate Loschmidt amplitude is needed on a large domain to obtain the work pdf by a Fourier transform. We studied here the sinh-Gordon model, differing from the free boson theory in a nontrivial scattering phase only, arguably the simplest case. So much so, that the continuation of the BTBA from the real positive half-axis (where it provides the partition function in finite volume and temperature) to the imaginary axis (providing the Loschmidt amplitude) is only nontrivial because of numerical obstacles. Namely, the evaluation of highly oscillatory integrals would be necessary. We proved a property of the BTBA equation by which it is possible to avoid such oscillatory integrals.

We obtained that by turning on the interaction the global properties of the pdf depart only slightly from the free result, however the details change. In particular the sequence of sharp edge singularities characteristic to free bosons turns into one with less pronounced peaks positioned near the energies of channels openings. The initial edge at the creation energy of two particles develops a fermionic singularity exponent.

\section*{Acknowledgments}
S.S. acknowledges financial support by SISSA -- International School for Advanced Studies under the ``Young SISSA Scientists Research Projects'' scheme 2011-2012 and by the ERC under Starting Grant 279391 EDEQS. T.P.thanks G. Mussardo for valuable discussions.

\end{document}